\documentclass[copyright]{eptcs}

\usepackage{epic}
\usepackage{eepic}
\usepackage{graphicx}
\usepackage{graphics}
\usepackage{amssymb}
\usepackage{amsmath}

\newcommand{\Tf}{\mbox{\bf f}}

\newcommand{\Ts}{\mbox{\bf s}}

\newcommand{\Tt}{\mbox{\bf t}}
\newcommand{\Tu}{\mbox{\bf u}}
\newcommand{\Tv}{\mbox{\bf v}}

\newcommand{\Tx}{\mbox{\bf x}}
\newcommand{\Ty}{\mbox{\bf y}}

\newcommand{\TT}{\mbox{\bf T}}

\newcommand{\bdisplay}{\begin{description}\footnotesize\item[]}
\newcommand{\edisplay}{\end{description}}

\newcommand{\bquot}[1]{\begin{quotation}\small\noindent
  \textbf{#1}\hspace{\labelsep}\ignorespaces}
\newcommand{\equot}{\unskip\end{quotation}}

\newcommand{\cG}{\mbox{$\cal G$}}

\newcommand{\txt}[1]{\mbox{ #1 }}

\newcommand{\noin}{\noindent}
\newcommand{\be}{\begin{enumerate}}
\newcommand{\ee}{\end{enumerate}}
\newcommand{\bi}{\begin{itemize}}
\newcommand{\ei}{\end{itemize}}

\newcommand{\ie}{{\it i.e.\/}} 

\newcommand{\etc}{{\it etc.\/}}

\newcommand{\olra}{\overleftrightarrow}

\newtheorem{theorem}{Theorem}
\newtheorem{definition}{Definition}
\newtheorem{lemma}{Lemma}
\newtheorem{example}{Example}
\newtheorem{prop}{Proposition}
\newtheorem{corollary}{Corollary}

\providecommand{\event}{DCFS 2010} 
\usepackage{breakurl}             

\title{On the Complexity of the Evaluation of Transient Extensions of Boolean Functions\thanks{This research was supported by the Natural Sciences and Engineering Research
Council of Canada under a grant 
and  a Postgraduate
Scholarship, and by a Graduate Award from the Department of Computer
Science, University of Toronto.}
}
\author{Janusz Brzozowski  \qquad\qquad Baiyu Li
\institute{David R.~Cheriton School of Computer Science\\
 University of Waterloo,
Waterloo, ON\\ Canada N2L 3G1}
 \email{\{brzozo, b5li\}@uwaterloo.ca}
\and
Yuli Ye
\institute{Department of Computer Science\\
 University of Toronto,
Toronto, ON\\
Canada M5S 3G4}
\email{y3ye@cs.toronto.edu}
}
\def\titlerunning{Complexity of the Evaluation of Transient Extensions}
\def\authorrunning{J. Brzozowski, B. Li \& Y. Ye}
\begin{document}
\maketitle

\begin{abstract}
Transient algebra is a multi-valued algebra for hazard detection in gate circuits. Sequences of alternating 0's and 1's, called transients, represent signal values, and gates are modeled by extensions of boolean functions to transients.
Formulas  for computing the output transient of a gate from the input transients are known for {\sc not, and, or} and {\sc xor} gates and their complements,
but, in general, even the problem of deciding whether the length of the output transient exceeds a given bound is NP-complete. We propose a method of evaluating extensions of general boolean functions. 
We introduce and study a class of functions with the following property: Instead of evaluating an extension of
a boolean function on a given set of transients, it is possible to get the
same value by using transients derived from the given ones, but
having length at most 3. 
We prove that all functions of three variables,  as well as certain other  functions, have this property, and can be efficiently evaluated.
\smallskip

\noin

\end{abstract}

\section{Introduction}

In 2003 Brzozowski and \'Esik~\cite{BrEs03} proposed an infinite algebra as a basis for a theory of hazards in gate circuits.
The fundamental concept  in this theory is that of a ``transient'', which is a nonempty alternating sequence of 0's and 1's representing a series of signal values. Boolean functions that are normally used to model gates are extended to transients. Given a boolean function $f(x_1,\ldots,x_n)$, and $n$ transients $\Tx_1,\ldots,\Tx_n$, the extension $\Tf(\Tx_1,\ldots,\Tx_n)$ of $f$ to transients is  defined as the longest transient that can be obtained by considering all possible orders of changes of the input variables.

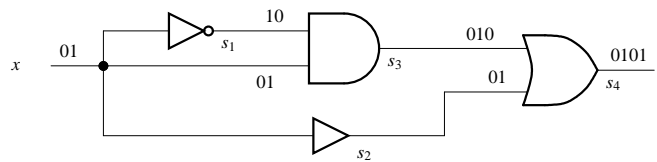
\begin{figure}[hbt]
\begin{center}
\setlength{\unitlength}{0.00030833in}
\begingroup\makeatletter\ifx\SetFigFont\undefined%
\gdef\SetFigFont#1#2#3#4#5{%
  \reset@font\fontsize{#1}{#2pt}%
  \fontfamily{#3}\fontseries{#4}\fontshape{#5}%
  \selectfont}%
\fi\endgroup%
{\renewcommand{\dashlinestretch}{30}
\begin{picture}(11052,2689)(0,-10)
\put(5940,109){\makebox(0,0)[lb]{\smash{{\SetFigFont{7}{8.4}{\familydefault}{\mddefault}{\updefault}$s_2$}}}}
\put(10215,1309){\makebox(0,0)[lb]{\smash{{\SetFigFont{7}{8.4}{\familydefault}{\mddefault}{\updefault}$s_4$}}}}
\put(6465,1684){\makebox(0,0)[lb]{\smash{{\SetFigFont{7}{8.4}{\familydefault}{\mddefault}{\updefault}$s_3$}}}}
\put(840,1834){\makebox(0,0)[lb]{\smash{{\SetFigFont{7}{8.4}{\familydefault}{\mddefault}{\updefault}$01$}}}}
\put(15,1609){\makebox(0,0)[lb]{\smash{{\SetFigFont{7}{8.4}{\familydefault}{\mddefault}{\updefault}$x$}}}}
\put(4365,2434){\makebox(0,0)[lb]{\smash{{\SetFigFont{7}{8.4}{\familydefault}{\mddefault}{\updefault}10}}}}
\put(7815,2134){\makebox(0,0)[lb]{\smash{{\SetFigFont{7}{8.4}{\familydefault}{\mddefault}{\updefault}010}}}}
\put(8190,1384){\makebox(0,0)[lb]{\smash{{\SetFigFont{7}{8.4}{\familydefault}{\mddefault}{\updefault}01}}}}
\put(10290,1759){\makebox(0,0)[lb]{\smash{{\SetFigFont{7}{8.4}{\familydefault}{\mddefault}{\updefault}0101}}}}
\put(3615,1984){\makebox(0,0)[lb]{\smash{{\SetFigFont{7}{8.4}{\familydefault}{\mddefault}{\updefault}$s_1$}}}}
\put(4215,1309){\makebox(0,0)[lb]{\smash{{\SetFigFont{7}{8.4}{\familydefault}{\mddefault}{\updefault}$01$}}}}
\put(1590,1684){\blacken\ellipse{150}{150}}
\put(1590,1684){\ellipse{150}{150}}
\path(1590,2284)(2715,2284)
\path(3465,2284)(5115,2284)
\path(5115,1684)(1590,1684)
\path(7440,1234)(8865,1234)
\path(6315,1984)(8865,1984)
\path(10065,1609)(11040,1609)
\path(1590,2284)(1590,1759)
\path(690,1684)(1440,1684)(1515,1684)
\path(5790,484)(7440,484)
\path(7440,1234)(7440,484)
\path(5190,484)(1590,484)(1590,1609)
\thicklines
\path(5190,784)(5190,184)(5790,484)(5190,784)
\put(3390,2284){\ellipse{150}{150}}
\path(2715,2584)(2715,1984)(3315,2284)(2715,2584)
\path(8790,1009)(9165,1009)
\path(8790,2209)(9165,2209)
\path(9165,2209)(9166,2209)(9169,2209)
	(9177,2209)(9191,2208)(9212,2207)
	(9237,2206)(9266,2205)(9297,2203)
	(9328,2200)(9359,2198)(9387,2195)
	(9413,2191)(9438,2188)(9460,2183)
	(9481,2178)(9501,2173)(9521,2166)
	(9540,2159)(9557,2152)(9575,2144)
	(9592,2135)(9610,2126)(9628,2115)
	(9647,2104)(9666,2092)(9685,2079)
	(9704,2066)(9722,2052)(9741,2038)
	(9759,2023)(9776,2009)(9793,1994)
	(9809,1979)(9824,1965)(9839,1951)
	(9852,1937)(9865,1923)(9878,1909)
	(9891,1894)(9903,1878)(9915,1862)
	(9927,1845)(9940,1827)(9952,1808)
	(9965,1787)(9979,1764)(9993,1740)
	(10007,1715)(10021,1690)(10034,1667)
	(10045,1646)(10054,1630)(10060,1619)
	(10063,1612)(10065,1609)
\path(9165,1009)(9166,1009)(9169,1009)
	(9177,1009)(9191,1010)(9212,1011)
	(9237,1012)(9266,1013)(9297,1015)
	(9328,1018)(9359,1020)(9387,1023)
	(9413,1027)(9438,1030)(9460,1035)
	(9481,1040)(9501,1045)(9521,1052)
	(9540,1059)(9557,1066)(9575,1074)
	(9592,1083)(9610,1092)(9628,1103)
	(9647,1114)(9666,1126)(9685,1139)
	(9704,1152)(9722,1166)(9741,1180)
	(9759,1195)(9776,1209)(9793,1224)
	(9809,1239)(9824,1253)(9839,1267)
	(9852,1281)(9865,1295)(9878,1309)
	(9891,1324)(9903,1340)(9915,1356)
	(9927,1373)(9940,1391)(9952,1410)
	(9965,1431)(9979,1454)(9993,1478)
	(10007,1503)(10021,1528)(10034,1551)
	(10045,1572)(10054,1588)(10060,1599)
	(10063,1606)(10065,1609)
\path(8790,2209)(8791,2206)(8793,2200)
	(8797,2190)(8802,2174)(8810,2152)
	(8819,2125)(8829,2094)(8840,2059)
	(8852,2022)(8864,1985)(8875,1948)
	(8886,1911)(8896,1876)(8905,1843)
	(8913,1811)(8919,1782)(8925,1754)
	(8930,1728)(8934,1703)(8937,1679)
	(8939,1655)(8940,1632)(8940,1609)
	(8940,1586)(8939,1563)(8937,1539)
	(8934,1515)(8930,1490)(8925,1464)
	(8919,1436)(8913,1407)(8905,1375)
	(8896,1342)(8886,1307)(8875,1270)
	(8864,1233)(8852,1196)(8840,1159)
	(8829,1124)(8819,1093)(8810,1066)
	(8802,1044)(8797,1028)(8793,1018)
	(8791,1012)(8790,1009)
\put(5715.000,1984.000){\arc{1200.000}{4.7124}{7.8540}}
\path(5715,2584)(5115,2584)(5115,1384)(5715,1384)
\end{picture}
}
\end{center}
\caption{Circuit with hazards.}
\label{fig:haz2}
\end{figure}

For example, consider the circuit of Fig.~\ref{fig:haz2}. 
A change in the input $x$ from 0 to 1 is represented  by the transient 01. The output of the inverter will then have the transient 10. If the lower input to the {\sc and} gate  changes before the upper input because of a delay in the inverter, the {\sc and} gate might have a hazard pulse denoted by the transient 010. This pulse is called a \emph{static hazard}; it is static because the {\sc and} gate's output is not supposed to change, but there may be an undesired transient pulse.
Now, if the {\sc or} gate's top input changes twice before its bottom input changes, we might have the transient 
0101 at the output of the {\sc or} gate. This is a \emph{dynamic hazard}, because the output of the {\sc or} gate is supposed to change; however, instead of changing just once it changes three times. 
Since the detection of hazards is an important problem in circuit analysis and design~\cite{BrEs03,BrSe95},  we are interested in evaluating extensions of boolean functions to predict the worst-case hazards, that is, the longest possible transients.

While the definition of the value of the extension of a boolean function is straightforward, it involves the construction of an $n$-dimensional directed graph in which all possible orders of input changes are shown. Since the size of this graph is exponential in the length of the input transients,  this method is inefficient.

There exist simple formulas~\cite{BrEs03} for  common boolean functions like {\sc not, and, or} and {\sc xor}, and
such formulas have been extended to any function obtained from the set $\{\mbox{\sc{or, xor}}\}$ by complementing any number of inputs, and/or the output~\cite{BrYe10}.
However, function composition does not preserve extensions~\cite{BrEs03}, and the evaluation problem remains open
for general  functions.
Functions that are more complex than {\sc or}, {\sc and}, {\sc xor}, {\sc nor}, {\sc nand} and {\sc xnor}
occur frequently in CMOS implementations~\cite{BrSe95}.
We study extensions of general boolean functions, and propose ways of evaluating them. In particular, we introduce a method in which an arbitrary vector $\Tx$ of transients is replaced by a vector $\tilde{\Tx}$ of ``characteristic transients" which are of length at most~3. We show that evaluating $\Tf(\Tx)$ can be reduced to evaluating $\Tf(\tilde{\Tx})$ for some functions. This makes it possible to efficiently evaluate all the functions of three or fewer variables, and some other functions with special properties.

The remainder of the paper is structured as follows. Transients,  vectors of transients, and extensions of boolean functions to transients are defined in Section~\ref{sec:trans}. The evaluation of functions in a certain class $\mathcal G$ is considered in Section~\ref{sec:G}, where we define the ``cost'' of a transient vector $\Tx$ to be the difference between the number of changes in $\Tx$  and the number of changes in $\Tf(\Tx)$. The concept of cost is extended to paths in digraphs and walks in boolean cubes in Section~\ref{sec:cubes}. 
\goodbreak
Characteristic vectors are defined in Section~\ref{sec:char}.
 In Section~\ref{sec:sym} we prove that all 3-variable functions can be efficiently evaluated using characteristic vectors, and that there exists a 5-variable function that cannot be so evaluated. Section~\ref{sec:conc} concludes the paper.

\section{Transients, vectors, and extensions of functions}
\label{sec:trans}
The cardinality of a set $S$ is  $|S|$. For $n\geq1$, let $[n]=\{1,\ldots,n\}$. If $A$ is an alphabet, then $A^*$ ($A^+$) denotes the free monoid (free semigroup) generated by~$A$. 
The length of a word $w\in A^*$ is $l(w)$,  and the first and last letters of $w\in A^+$ are $\alpha(w)$ and $\omega(w)$, respectively.
For boolean operations, we use $x'$ for complement, $xy$ for {\sc and},  $x+y$ for {\sc or}, and $x\oplus y$ for {\sc xor} (exclusive {\sc or}).

Let $B=\{0,1\}$; a {\em binary word\/} is any word in $B^*$. 
A {\em transient\/} is a binary word in $B^+$ of alternating 0's and 1's; thus the set $\TT$ of all transients is $0(10)^*+(01)^*01+(10)^*10+(10)^*1$, in regular-expression notation.
Transients are denoted by boldface letters.
A transient can be obtained from any nonempty binary word by
{\em contraction\/}, \ie, elimination of all duplicates immediately
following a symbol; thus contraction is a function from $B^+$ to $\TT$.
We denote the  contraction of a word $w$ by $\olra{w}$.
For example,  
$\olra{001000}=010$.
For $\Ts,\Tt\in \TT$,  ${\Ts} \circ {\Tt}$ is concatenation followed by contraction, \ie,
${\Ts} \circ {\Tt} = \olra{\Ts\Tt}$.
The $\circ$ operation is associative.

If $\Tt = t_1 \cdots t_m$ is a transient,  $t_i \in B$ for $i\in[m]$, then  $\Delta(\Tt) = l(\Tt) - 1 = m - 1$ is the {\em number of 
changes} in $\Tt$. 
A transient $\Tt$ is completely determined by its beginning $\alpha(\Tt)$ and the number of changes $\Delta(\Tt)$; thus we have another representation of $\Tt$ which we indicate by angle brackets: 
$ \Tt = t_1 \cdots t_m = \langle\alpha(\Tt);\Delta(\Tt)\rangle=
 \langle t_1;m-1\rangle. $
The number of 0's in a transient $\Tt$ is $z(\Tt)$, and the number of 1's (``units'') is  $u(\Tt)$.

A {\em prefix\/} of a transient $\Tt=t_1\cdots t_m$ is any transient $\Tu=t_1\cdots t_i$, where $i\in[m]$.
A {\em suffix\/} of a transient is defined similarly.
Note that we do not allow the empty word to be a prefix or suffix, because the empty word is not a transient. However, $\Tt$ \emph{is} a prefix and suffix of itself.
If $\Tu$ is a prefix of $\Tt$ and $l(\Tu)<l(\Tt)$, then there exists a transient $\Tv$, a suffix of $\Tt$, such that $\Tt=\Tu\Tv$.
A transient $\Ts$ is the {\em successor\/} of a transient  $\Tt=t_1\cdots t_m$ if and only if $\Ts=t_1\cdots t_mt_{m+1}$, where $t_i\in B$, for $i\in[m+1]$.
For example, $010$ is the successor of $01$.

A {\em transient vector,\/}   or simply a {\em vector,\/} is a tuple $\Tx=(\Tx_1,\ldots, \Tx_n)\in \TT^n$. 
By convention, if $\Tx$ is a vector, then $\Tx_i$ is a component of $\Tx$.
The $\circ$ operation is extended to transient vectors component-wise.
The {\em length\/} of a vector $\Tx=(\Tx_1,\ldots, \Tx_n)$ is $l(\Tx) = \sum_{i=1}^n l(\Tx_i)$.
The {\em number of changes\/} of $\Tx$ is $\Delta(\Tx) = \sum_{i=1}^n \Delta(\Tx_i) = l(\Tx) - n$. 
We also define vectors $\alpha(\Tx)=(\alpha(\Tx_1),\ldots, \alpha(\Tx_n))$, and $\omega(\Tx)=(\omega(\Tx_1),\ldots, \omega(\Tx_n))$. 
A vector is completely determined by its beginning $\alpha(\Tx)$ and the number of changes $\Delta(\Tx_i)$ of each component of $\Tx$; thus we have another representation  
 $ \Tx = \langle\alpha(\Tx);\Delta(\Tx_1),\ldots,\Delta(\Tx_n)\rangle. $
A vector $\Tu=(\Tu_1,\ldots, \Tu_n)$ is a {\em prefix\/} ({\em suffix}) of 
vector $\Tv=(\Tv_1,\ldots, \Tv_n)$ if $\Tu_i$ is a prefix (suffix) of $\Tv_i$ for all $i\in [n]$.
A vector  $\Tu=(\Tu_1,\ldots, \Tu_n)$ is a {\em successor\/}  of  $\Tv=(\Tv_1,\ldots, \Tv_n)$ if $\Tu_i$ is the successor of $\Tv_i$ for some $i\in [n]$ and $\Tu_j=\Tv_j$, for all $j\not=i$.
\goodbreak

Our terminology on graphs is from~\cite{BoMu76}. 
If $f:B^n\to B$ is a boolean function and
$\Tx=(\Tx_1,\ldots, \Tx_n)\in \TT^n$ is a  vector, we construct the {\em transient digraph\/} $D=D_f(\Tx)=(V,E,\psi,\lambda)$ of $f$ for $\Tx$, where $(V,E,\psi)$ is a digraph, $V$ (the set of \emph{vertices}) is the set of all prefixes of $\Tx$, $E$ (the set of \emph{arcs}) is $E=\{e=(\Tu,\Tv)\mid \Tv \txt{ is a successor of} \Tu\}$, 
$\psi$ (the \emph{incidence function} assigning to each arc of $D$ an ordered pair of vertices of~$D$) is 
$\psi(e)=\psi((\Tu,\Tv))=(\Tu,\Tv)$, and $\lambda:V\to B$ is the \emph{output function} assigning the value $f(\omega(\Tv))$ to every $\Tv\in V$. Each directed path 
$P= \Tv_1,\ldots,\Tv_m$ in $D$ from  $\Tv_1=\alpha(\Tx)$ to   $\Tv_m=\Tx$ has length $m=\sum_{i=1}^n \Delta(\Tx_i)$. 
We extend $\lambda$ to paths:  $\lambda(P) = \lambda(\Tv^1)\cdots\lambda(\Tv^m)$. 
Paths are always  from $\alpha(\Tx)$ to $\Tx$.
\begin{definition}
Let $f(x):B^n\to B$ be a boolean function. The {\em transient extension\/} (or simply {\em extension\/}) of $f$ is a function $\Tf(\Tx):\TT^n\to \TT$,
such that for any $\Tx=(\Tx_1,\ldots, \Tx_n)\in \TT^n$, 
$\Tf(\Tx) = \olra{\lambda({P})}$, where ${P}$ is a path in $D_f(\Tx)$ and  $\olra{\lambda({P})}$ is of maximal length; we call such a path ${P}$ {\em optimal}. 
\end{definition}

\begin{figure}[hbt]
\begin{center}
\setlength{\unitlength}{0.00050000in}
\begingroup\makeatletter\ifx\SetFigFont\undefined%
\gdef\SetFigFont#1#2#3#4#5{%
  \reset@font\fontsize{#1}{#2pt}%
  \fontfamily{#3}\fontseries{#4}\fontshape{#5}%
  \selectfont}%
\fi\endgroup%
{\renewcommand{\dashlinestretch}{30}
\begin{picture}(5074,3805)(0,-10)
\put(863,2144){\makebox(0,0)[lb]{\smash{{\SetFigFont{7}{8.4}{\familydefault}{\mddefault}{\updefault}$\Tv^3$}}}}
\put(817,2473){\arc{210}{1.5708}{3.1416}}
\put(817,2563){\arc{210}{3.1416}{4.7124}}
\put(1657,2563){\arc{210}{4.7124}{6.2832}}
\put(1657,2473){\arc{210}{0}{1.5708}}
\path(712,2473)(712,2563)
\path(817,2668)(1657,2668)
\path(1762,2563)(1762,2473)
\path(1657,2368)(817,2368)
\put(817,1573){\arc{210}{1.5708}{3.1416}}
\put(817,1663){\arc{210}{3.1416}{4.7124}}
\put(1657,1663){\arc{210}{4.7124}{6.2832}}
\put(1657,1573){\arc{210}{0}{1.5708}}
\path(712,1573)(712,1663)
\path(817,1768)(1657,1768)
\path(1762,1663)(1762,1573)
\path(1657,1468)(817,1468)
\put(817,673){\arc{210}{1.5708}{3.1416}}
\put(817,763){\arc{210}{3.1416}{4.7124}}
\put(1657,763){\arc{210}{4.7124}{6.2832}}
\put(1657,673){\arc{210}{0}{1.5708}}
\path(712,673)(712,763)
\path(817,868)(1657,868)
\path(1762,763)(1762,673)
\path(1657,568)(817,568)
\put(2467,3373){\arc{210}{1.5708}{3.1416}}
\put(2467,3463){\arc{210}{3.1416}{4.7124}}
\put(3307,3463){\arc{210}{4.7124}{6.2832}}
\put(3307,3373){\arc{210}{0}{1.5708}}
\path(2362,3373)(2362,3463)
\path(2467,3568)(3307,3568)
\path(3412,3463)(3412,3373)
\path(3307,3268)(2467,3268)
\put(4117,3373){\arc{210}{1.5708}{3.1416}}
\put(4117,3463){\arc{210}{3.1416}{4.7124}}
\put(4957,3463){\arc{210}{4.7124}{6.2832}}
\put(4957,3373){\arc{210}{0}{1.5708}}
\path(4012,3373)(4012,3463)
\path(4117,3568)(4957,3568)
\path(5062,3463)(5062,3373)
\path(4957,3268)(4117,3268)
\put(2467,2473){\arc{210}{1.5708}{3.1416}}
\put(2467,2563){\arc{210}{3.1416}{4.7124}}
\put(3307,2563){\arc{210}{4.7124}{6.2832}}
\put(3307,2473){\arc{210}{0}{1.5708}}
\path(2362,2473)(2362,2563)
\path(2467,2668)(3307,2668)
\path(3412,2563)(3412,2473)
\path(3307,2368)(2467,2368)
\put(4117,2473){\arc{210}{1.5708}{3.1416}}
\put(4117,2563){\arc{210}{3.1416}{4.7124}}
\put(4957,2563){\arc{210}{4.7124}{6.2832}}
\put(4957,2473){\arc{210}{0}{1.5708}}
\path(4012,2473)(4012,2563)
\path(4117,2668)(4957,2668)
\path(5062,2563)(5062,2473)
\path(4957,2368)(4117,2368)
\put(2467,1573){\arc{210}{1.5708}{3.1416}}
\put(2467,1663){\arc{210}{3.1416}{4.7124}}
\put(3307,1663){\arc{210}{4.7124}{6.2832}}
\put(3307,1573){\arc{210}{0}{1.5708}}
\path(2362,1573)(2362,1663)
\path(2467,1768)(3307,1768)
\path(3412,1663)(3412,1573)
\path(3307,1468)(2467,1468)
\put(2467,673){\arc{210}{1.5708}{3.1416}}
\put(2467,763){\arc{210}{3.1416}{4.7124}}
\put(3307,763){\arc{210}{4.7124}{6.2832}}
\put(3307,673){\arc{210}{0}{1.5708}}
\path(2362,673)(2362,763)
\path(2467,868)(3307,868)
\path(3412,763)(3412,673)
\path(3307,568)(2467,568)
\put(4117,1573){\arc{210}{1.5708}{3.1416}}
\put(4117,1663){\arc{210}{3.1416}{4.7124}}
\put(4957,1663){\arc{210}{4.7124}{6.2832}}
\put(4957,1573){\arc{210}{0}{1.5708}}
\path(4012,1573)(4012,1663)
\path(4117,1768)(4957,1768)
\path(5062,1663)(5062,1573)
\path(4957,1468)(4117,1468)
\put(4117,673){\arc{210}{1.5708}{3.1416}}
\put(4117,763){\arc{210}{3.1416}{4.7124}}
\put(4957,763){\arc{210}{4.7124}{6.2832}}
\put(4957,673){\arc{210}{0}{1.5708}}
\path(4012,673)(4012,763)
\path(4117,868)(4957,868)
\path(5062,763)(5062,673)
\path(4957,568)(4117,568)
\path(1762,3418)(2362,3418)
\blacken\path(2242.000,3388.000)(2362.000,3418.000)(2242.000,3448.000)(2242.000,3388.000)
\path(1762,2518)(2362,2518)
\blacken\path(2242.000,2488.000)(2362.000,2518.000)(2242.000,2548.000)(2242.000,2488.000)
\path(1762,1618)(2362,1618)
\blacken\path(2242.000,1588.000)(2362.000,1618.000)(2242.000,1648.000)(2242.000,1588.000)
\path(1762,718)(2362,718)
\blacken\path(2242.000,688.000)(2362.000,718.000)(2242.000,748.000)(2242.000,688.000)
\path(3412,718)(4012,718)
\blacken\path(3892.000,688.000)(4012.000,718.000)(3892.000,748.000)(3892.000,688.000)
\path(3412,1618)(4012,1618)
\blacken\path(3892.000,1588.000)(4012.000,1618.000)(3892.000,1648.000)(3892.000,1588.000)
\path(3412,2518)(4012,2518)
\blacken\path(3892.000,2488.000)(4012.000,2518.000)(3892.000,2548.000)(3892.000,2488.000)
\path(3412,3418)(4012,3418)
\blacken\path(3892.000,3388.000)(4012.000,3418.000)(3892.000,3448.000)(3892.000,3388.000)
\path(1237,2668)(1237,3268)
\blacken\path(1267.000,3148.000)(1237.000,3268.000)(1207.000,3148.000)(1267.000,3148.000)
\path(1237,1768)(1237,2368)
\blacken\path(1267.000,2248.000)(1237.000,2368.000)(1207.000,2248.000)(1267.000,2248.000)
\path(1237,868)(1237,1468)
\blacken\path(1267.000,1348.000)(1237.000,1468.000)(1207.000,1348.000)(1267.000,1348.000)
\path(2887,868)(2887,1468)
\blacken\path(2917.000,1348.000)(2887.000,1468.000)(2857.000,1348.000)(2917.000,1348.000)
\path(2887,2668)(2887,3268)
\blacken\path(2917.000,3148.000)(2887.000,3268.000)(2857.000,3148.000)(2917.000,3148.000)
\path(4537,2668)(4537,3268)
\blacken\path(4567.000,3148.000)(4537.000,3268.000)(4507.000,3148.000)(4567.000,3148.000)
\path(4537,1768)(4537,2368)
\blacken\path(4567.000,2248.000)(4537.000,2368.000)(4507.000,2248.000)(4567.000,2248.000)
\path(4537,868)(4537,1468)
\blacken\path(4567.000,1348.000)(4537.000,1468.000)(4507.000,1348.000)(4567.000,1348.000)
\path(2887,1768)(2887,2368)
\blacken\path(2917.000,2248.000)(2887.000,2368.000)(2857.000,2248.000)(2917.000,2248.000)
\path(134,118)(884,118)
\blacken\path(764.000,88.000)(884.000,118.000)(764.000,148.000)(764.000,88.000)
\path(135,126)(135,876)
\blacken\path(165.000,756.000)(135.000,876.000)(105.000,756.000)(165.000,756.000)
\put(1312,2743){\makebox(0,0)[lb]{\smash{{\SetFigFont{7}{8.4}{\familydefault}{\mddefault}{\updefault}$0$}}}}
\put(4612,2743){\makebox(0,0)[lb]{\smash{{\SetFigFont{7}{8.4}{\familydefault}{\mddefault}{\updefault}$0$}}}}
\put(1312,3643){\makebox(0,0)[lb]{\smash{{\SetFigFont{7}{8.4}{\familydefault}{\mddefault}{\updefault}$1$}}}}
\put(2887,3643){\makebox(0,0)[lb]{\smash{{\SetFigFont{7}{8.4}{\familydefault}{\mddefault}{\updefault}$1$}}}}
\put(2962,2743){\makebox(0,0)[lb]{\smash{{\SetFigFont{7}{8.4}{\familydefault}{\mddefault}{\updefault}$1$}}}}
\put(2962,1843){\makebox(0,0)[lb]{\smash{{\SetFigFont{7}{8.4}{\familydefault}{\mddefault}{\updefault}$1$}}}}
\put(4612,1843){\makebox(0,0)[lb]{\smash{{\SetFigFont{7}{8.4}{\familydefault}{\mddefault}{\updefault}$1$}}}}
\put(1312,1843){\makebox(0,0)[lb]{\smash{{\SetFigFont{7}{8.4}{\familydefault}{\mddefault}{\updefault}$1$}}}}
\put(2962,943){\makebox(0,0)[lb]{\smash{{\SetFigFont{7}{8.4}{\familydefault}{\mddefault}{\updefault}$1$}}}}
\put(1312,943){\makebox(0,0)[lb]{\smash{{\SetFigFont{7}{8.4}{\familydefault}{\mddefault}{\updefault}$0$}}}}
\put(4612,943){\makebox(0,0)[lb]{\smash{{\SetFigFont{7}{8.4}{\familydefault}{\mddefault}{\updefault}$0$}}}}
\put(4612,3643){\makebox(0,0)[lb]{\smash{{\SetFigFont{7}{8.4}{\familydefault}{\mddefault}{\updefault}$1$}}}}
\put(862,3365){\makebox(0,0)[lb]{\smash{{\SetFigFont{7}{8.4}{\familydefault}{\mddefault}{\updefault}$(0,1010)$}}}}
\put(4056,3358){\makebox(0,0)[lb]{\smash{{\SetFigFont{7}{8.4}{\familydefault}{\mddefault}{\updefault}$(010,1010)$}}}}
\put(2475,3359){\makebox(0,0)[lb]{\smash{{\SetFigFont{7}{8.4}{\familydefault}{\mddefault}{\updefault}$(01,1010)$}}}}
\put(915,2458){\makebox(0,0)[lb]{\smash{{\SetFigFont{7}{8.4}{\familydefault}{\mddefault}{\updefault}$(0,101)$}}}}
\put(2520,2450){\makebox(0,0)[lb]{\smash{{\SetFigFont{7}{8.4}{\familydefault}{\mddefault}{\updefault}$(01,101)$}}}}
\put(4117,2458){\makebox(0,0)[lb]{\smash{{\SetFigFont{7}{8.4}{\familydefault}{\mddefault}{\updefault}$(010,101)$}}}}
\put(4169,1551){\makebox(0,0)[lb]{\smash{{\SetFigFont{7}{8.4}{\familydefault}{\mddefault}{\updefault}$(010,10)$}}}}
\put(2572,1551){\makebox(0,0)[lb]{\smash{{\SetFigFont{7}{8.4}{\familydefault}{\mddefault}{\updefault}$(01,10)$}}}}
\put(974,1550){\makebox(0,0)[lb]{\smash{{\SetFigFont{7}{8.4}{\familydefault}{\mddefault}{\updefault}$(0,10)$}}}}
\put(997,666){\makebox(0,0)[lb]{\smash{{\SetFigFont{7}{8.4}{\familydefault}{\mddefault}{\updefault}$(0,1)$}}}}
\put(2611,659){\makebox(0,0)[lb]{\smash{{\SetFigFont{7}{8.4}{\familydefault}{\mddefault}{\updefault}$(01,1)$}}}}
\put(4237,658){\makebox(0,0)[lb]{\smash{{\SetFigFont{7}{8.4}{\familydefault}{\mddefault}{\updefault}$(010,1)$}}}}
\put(2497,3051){\makebox(0,0)[lb]{\smash{{\SetFigFont{7}{8.4}{\familydefault}{\mddefault}{\updefault}$\Tv^5$}}}}
\put(4155,3058){\makebox(0,0)[lb]{\smash{{\SetFigFont{7}{8.4}{\familydefault}{\mddefault}{\updefault}$\Tv^6$}}}}
\put(2512,2158){\makebox(0,0)[lb]{\smash{{\SetFigFont{7}{8.4}{\familydefault}{\mddefault}{\updefault}$\Tv^4$}}}}
\put(15,935){\makebox(0,0)[lb]{\smash{{\SetFigFont{7}{8.4}{\familydefault}{\mddefault}{\updefault}$x_2$}}}}
\put(975,73){\makebox(0,0)[lb]{\smash{{\SetFigFont{7}{8.4}{\familydefault}{\mddefault}{\updefault}$x_1$}}}}
\put(847,367){\makebox(0,0)[lb]{\smash{{\SetFigFont{7}{8.4}{\familydefault}{\mddefault}{\updefault}$\Tv^1$}}}}
\put(855,1243){\makebox(0,0)[lb]{\smash{{\SetFigFont{7}{8.4}{\familydefault}{\mddefault}{\updefault}$\Tv^2$}}}}
\put(817,3373){\arc{210}{1.5708}{3.1416}}
\put(817,3463){\arc{210}{3.1416}{4.7124}}
\put(1657,3463){\arc{210}{4.7124}{6.2832}}
\put(1657,3373){\arc{210}{0}{1.5708}}
\path(712,3373)(712,3463)
\path(817,3568)(1657,3568)
\path(1762,3463)(1762,3373)
\path(1657,3268)(817,3268)
\end{picture}
}
\end{center}
\caption{Digraph $D_f(010,1010)$ for $f=x_1+{x'_2}$.}
\label{extension}
\end{figure}
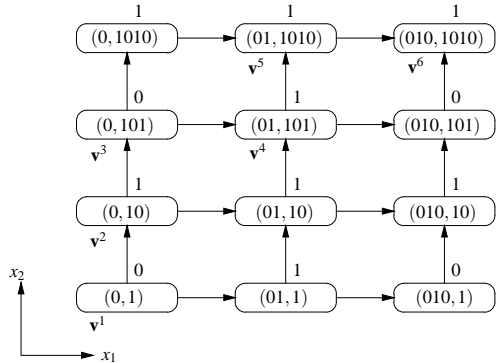

\begin{example}
In Fig.~\ref{extension} we show the digraph $D_f(010,1010)$ for $f=x_1+x'_2$, where changes in $x_1$ are  horizontal, and  in $x_2$, vertical. The initial vertex is $\alpha(010,1010)=(0,1)$. 
If the inputs are changed in the order $x_2,x_2,x_1,x_2,x_1$ (path $\Tv^1,\Tv^2,\ldots,\Tv^6$), then the binary word defined by $\lambda$ is $010111$, and its contraction
is $0101$. The longest output is $(01)^3$, corresponding to the optimal path in which the changes are made in the order $x_1,x_1,x_2,x_2,x_2$, and so $\Tf(010,1010)=(01)^3$.
\end{example}

For vector $\Tx \in \TT^n$, let $\varphi(\Tx)$ be the number of paths in $D_f(\Tx)$, let $m = \Delta(\Tx)$, and let $m_i = \Delta(\Tx_i)$ for $i \in [n]$. Then
\goodbreak
\begin{equation}
\varphi(\Tx) = {m \choose m_1,\ldots,m_n} = \frac{m!}{m_1!\cdots m_n!};
\end{equation}
that is, $\varphi(\Tx)$ is a multinomial coefficient. 
The maximal value of $\varphi(\Tx)$ has the following approximation~\cite{RiSh08}: 
\begin{eqnarray}
\label{eq:phi}
  \varphi(\Tx) \approx (2\pi m)^{\frac{1-n}{2}}n^{m+\frac{n}{2}}.
\end{eqnarray}
We usually consider $n$ to be small or fixed; then $\varphi(\Tx)$ is exponential in $m$.
Consequently, the obvious way to evaluate $\Tf(\Tx)$ is not feasible because of the large number of paths to explore. 

\section{Functions in class $\cal G$}
\label{sec:G}
In contrast to the general case above, for {\sc not, xor, or} and {\sc and} there are simple formulas~\cite{BrEs03}:
If $\Tt=t_1\cdots t_m$, then
\begin{equation} 
\label{eq:comp}
\Tt'=(t_1\cdots t_m)'=t'_1\cdots t'_m.
\end{equation}
If   $f(x_1,\ldots,x_n) = x_1 \oplus \cdots \oplus x_n$ is   {\sc xor},  then, for all $\Tx=(\Tx_1,\ldots,\Tx_n) \in \TT^n$,
\begin{eqnarray}
\label{axor}
\alpha(\Tf(\Tx)) &=& \alpha(\Tx_1) \oplus \cdots \oplus \alpha(\Tx_n),\quad \omega(\Tf(\Tx)) =\omega(\Tx_1) \oplus \cdots \oplus \omega(\Tx_n),\\
\label{lxor}
l(\Tf(\Tx)) &=& 1+  \sum_{i=1}^n (l(\Tx_i)-1).
\end{eqnarray} 
If  $f(x_1,\ldots,x_n) = x_1 + \cdots + x_n$ is {\sc or}, then,
for all $\Tx=(\Tx_1,\ldots,\Tx_n) \in \TT^n$,
\begin{eqnarray}
\label{aor}
\alpha(\Tf(\Tx)) &=& \alpha(\Tx_1) + \cdots + \alpha(\Tx_n),\quad \omega(\Tf(\Tx)) = \omega(\Tx_1) + \cdots + \omega(\Tx_n),\\
z(\Tf(\Tx)) &=& 
\label{zor}
\left\{
\begin{array}{ll}
0, & \txt{if} \ \exists i\in [n] \ \Tx_i = 1;\\
1 + \sum_{i=1}^n (z(\Tx_i)-1), & \txt{otherwise.} 
\end{array}
\right. 
\end{eqnarray} 
If  $f(x_1,\ldots,x_n) = x_1  \cdots  x_n$ is {\sc and}, then, 
for all $\Tx=(\Tx_1,\ldots,\Tx_n) \in \TT^n$,
\begin{eqnarray}
\alpha(\Tf(\Tx)) &=& \alpha(\Tx_1)  \cdots  \alpha(\Tx_n),\quad \omega(\Tf(\Tx)) = \omega(\Tx_1)  \cdots  \omega(\Tx_n),\\
u(\Tf(\Tx)) &=& 
\left\{
\begin{array}{ll}
0, & \txt{if} \ \exists i\in [n] \ \Tx_i = 0;\\
1 + \sum_{i=1}^n (u(\Tx_i)-1), & \txt{otherwise.} 
\end{array}
\right. 
\end{eqnarray} 
Using these formulas we can evaluate transient extensions of {\sc not, xor, or} and {\sc and} in the time linear in the length of the input vector. For example, to evaluate {\sc or} for $\Tx \in \TT^n$, we compute $\alpha(\Tf(\Tx))$, $\omega(\Tf(\Tx))$, and  the number of 0's in $\Tx$. 

The class  ${\mathcal G}$ of boolean functions is defined as follows~\cite{BrYe10}:
\begin{definition}
Let ${\mathcal H}=\{${\sc or, xor}$\}$ and let ${\mathcal G}$
be the set of functions obtained by complementing any number of inputs and/or the output of functions from ${\mathcal H}$;
here {\sc or} and {\sc xor}  may have  any non-zero number  of inputs, including one.  
\end{definition}
Note that a 1-input {\sc or} or {\sc xor} function is the identity function, and that $\mathcal G$ includes all the boolean functions of two variables, except the constants 0 and 1, as well as  {\sc and, nor, nand} and {\sc xnor} functions with any numbers of inputs.

It was proved in~\cite{BrYe10} that functions in $\cG$ can be evaluated by complementing the input transients of any complemented arguments, and by complementing the output transient, if the function itself is complemented. 
Consequently, we have
\begin{prop}
\label{prop:funcG}
Functions in $\cG$ can be evaluated in the time linear in the length of the input vector. 
\end{prop}
For example, if $f(x_1,x_2)=(x_1+x'_2)'$, then
$
\Tf(010,10)=(010+(10)')'=(010+01)'=(0101)'=1010.
$
However, in general, function composition does not preserve extensions~\cite{BrEs03}. For example, by~(\ref{axor}) and (\ref{lxor}), 
$01\oplus 101=1010$, but  if we express $\Ts\oplus\Tt$ as $\Ts\Tt'+\Ts'\Tt$,  we get
$101010$. For this reason, we need to consider functions that are not in $\cG$ separately.

As we have seen, evaluating a transient extension  from the transient digraph is not efficient. In \cite{BrEs03} it is shown that even the problem of estimating the length of $\Tf(\Tx)$ is NP-complete. However, the concept of ``cost" that we are about to define makes the calculation feasible for some functions.

\begin{definition}
 Let $f : B^n \rightarrow B$ be a boolean function, and $\Tf : \TT^n \rightarrow \TT$, its extension. Let $\Tx = (\Tx_1, \ldots, \Tx_2) = \langle \alpha(\Tx);\Delta(\Tx_1),\ldots,\Delta(\Tx_n) \rangle$ be a transient vector. 
 The {\em cost of $\Tx$ for $f$} is
$ c_f(\Tx) = \Delta(\Tx) - \Delta(\Tf(\Tx)) = \sum_{i=1}^n \Delta(\Tx_i) - \Delta(\Tf(\Tx)). $
\end{definition}

The following upper bound for $l(\Tf(\Tx))$ is given in~\cite{BrEs03}: 
\begin{equation}
l(\Tf(\Tx)) \leq 1 + \sum_{i=1}^n(l(\Tx_i)-1) = 1 + \sum_{i=1}^n \Delta(\Tx_i) = 1 + \Delta(\Tx).
\end{equation}
Thus $\Delta(\Tf(\Tx)) = l(\Tf(\Tx)) - 1 \leq \Delta(\Tx)$, and so $c_f(\Tx)$ is a non-negative integer.
If we know $\Tx$ and its cost $c_f(\Tx)$, then we can easily evaluate $\Tf(\Tx)$ as follows:
\begin{equation}
\Tf(\Tx)=\langle \alpha(\Tf(\Tx)); \Delta(\Tf(\Tx))\rangle=
\langle \alpha(\Tf(\Tx)); \Delta(\Tx)-c_f(\Tx)\rangle .
\end{equation}
\begin{example}
For $\Tx = (0101,10101)= \langle (0,1),3,4 \rangle$, 
 $ \Delta(\Tx)= 7. $
 For $f(x) = x_1 + x_2'$, using (\ref{eq:comp}), (\ref{aor}) and (\ref{zor}), we have  
 $\Tf(\Tx) = (01)^4$, $\Delta(\Tf(\Tx))=7$,
  and 
 $ c_f(\Tx) =  7-7=  0$.
For  $\Ty = (0101,0101010) = \langle (0,0), 3, 6 \rangle$,
if $f(y) = y_1 + y_2'$, we have
 $ \Tf(\Ty) = (10)^41$, 
  and  $c_f(\Ty) =9-8=1$.
\end{example}

A binary vector $(x_1,\ldots,x_n)$ with $x_i=0$, for all $i\in [n]$ is denoted by $0^n$. 
We define {\em non-negative subtraction\/} $m \ominus n$ of integer $n$ from integer $m$ as 
$m\ominus n=m-n$ if $m\ge n$, and $m\ominus n=0$, otherwise.
A transient $\Tt=t_1\cdots t_m$ is {\em proper\/} if its length is at least 2, \ie, if it contains at least one change.
A vector is {\em proper\/} if all of its components are proper.

\begin{theorem}
\label{thm:costg}
If $\Tx$ is proper, then
\begin{eqnarray}
\label{eq:costxor}
& c_{xor}(\Tx) = 0, \\
\label{eq:costor}
& c_{or}(\Tx)=(u(\alpha(\Tx))\ominus 1) + (u(\omega(\Tx))\ominus 1), \\
\label{eq:costand}
& c_{and}(\Tx)=(z(\alpha(\Tx))\ominus 1) + (z(\omega(\Tx))\ominus 1). 
\end{eqnarray}
\end{theorem}

\noin{\it Proof:\/}
If  $f$ is {\sc xor},  we have
$c_{xor} = \Delta(\Tx) - \Delta(\Tf(\Tx)) = 
\sum_{i=1}^n \Delta(\Tx_i) - \Delta(\Tf(\Tx))=
\sum_{i=1}^n (l(\Tx_i)-1) - (l(\Tx) - 1) = 0,$
where we have used Equation~(\ref{lxor}).
\goodbreak

If  $f$ is {\sc or}, we consider three cases:

\begin{enumerate}
  \item \label{case1} $u(\alpha(\Tx)) = 0$.
    We have  $u(\alpha(\Tx)) \ominus 1 = 0$. Since $\alpha(\Tx_i) = 0$ for all $i \in [n]$, we have
    $\alpha(f(\Tx)) = 0$.
    Let $S = \{i \mid \omega(\Tx_i) = 1\}$, $T = \{i \mid \omega(\Tx_i) = 0\}$. Then
    \begin{displaymath}
      \begin{array}{rl}
        \Delta(\Tx) &= \sum_{i=1}^n (l(\Tx_i) - 1) = \sum_{i \in S} (l(\Tx_i) - 1) + \sum_{i \in T} (l(\Tx_i) - 1)\\ 
        &= \sum_{i \in S} (2z(\Tx_i) - 1) + \sum_{i \in T} (2z(\Tx_i) - 2) \\
        &= \sum_{i=1}^n (2z(\Tx_i) - 2) + |S| =2(z(f(\Tx)) - 1) + u(\omega(\Tx)),
      \end{array}
    \end{displaymath}
where the last equality uses Equation~(\ref{zor}).
    If $u(\omega(\Tx)) = 0$, then $\omega(\Tx_i) = 0$ for all $i \in [n]$, and $\omega(f(\Tx)) = 0$. Thus we have $l(f(\Tx)) = 2z(f(\Tx)) - 1$, and
    $\Delta(\Tx) = l(f(\Tx)) - 1 = \Delta(\Tf(\Tx)). $
    Then $c_{or}(\Tx) = \Delta(\Tx) - \Delta(\Tf(\Tx)) = 0 = (u(\alpha(\Tx))\ominus 1) + (u(\omega(\Tx))\ominus 1)$.
    Otherwise, $u(\omega(\Tx)) \geq 1$. Then $\omega(f(\Tx)) = 1$, and $l(f(\Tx)) = 2z(f(\Tx))$. Thus
    \begin{displaymath}
      \begin{array}{rl}
        \Delta(\Tx) 
        &= 2(z(f(\Tx)) - 1) +u(\omega(\Tx)) = l(f(\Tx)) - 1 + u(\omega(\Tx)) - 1\\
        &= l(f(\Tx)) - 1 + u(\omega(\Tx)) \ominus 1 = \Delta(\Tf(\Tx)) + u(\omega(\Tx)) \ominus 1.\\
      \end{array}
    \end{displaymath}
    Hence, $c_{or}(\Tx) = 
    u(\omega(\Tx)) \ominus 1 = (u(\alpha(\Tx))\ominus 1) + (u(\omega(\Tx))\ominus 1)$.

  \item \label{case2} $u(\omega(\Tx)) = 0$. This case is symmetric to Case~\ref{case1}. 

  \item $u(\alpha(\Tx)) \neq 0$, and $u(\omega(\Tx)) \neq 0$.
    Since $\Tx$ is proper, then for all $i \in [n]$, we have $l(\Tx_i) \geq 2$, and there is at least one 0 in $\Tx_i$. Let $\Tx_i = \Tv_i \circ \Tu_i$, where $\omega(\Tv_i) = \alpha(\Tu_i) = 0$, $\Tv = (\Tv_1,\ldots,\Tv_n)$, and $\Tu = (\Tu_1,\ldots,\Tu_n)$. Then $\Tx = \Tv \circ \Tu$, and $\omega(\Tv) = \alpha(\Tu) = (0,\ldots,0) = 0^n$. By Cases~\ref{case1} and~\ref{case2},
    $ c_{or}(\Tv) = u(\alpha(\Tx)) \ominus 1, \quad \mathrm{and} \quad \quad c_{or}(\Tu) = u(\omega(\Tx)) \ominus 1. $
    Therefore, 
$c_{or}(\Tx) = c_{or}(\Tv) + c_{or}(\Tu) 
                 = (u(\alpha(\Tx)) \ominus 1) + (u(\omega(\Tx)) \ominus 1).
                 $

\end{enumerate}

By {\sc and}/{\sc or} duality, we have
$c_{and}(\Tx)=(z(\alpha(\Tx))\ominus 1) + (z(\omega(\Tx))\ominus 1).$

\section{Costs of paths in digraphs and walks in cubes}
\label{sec:cubes}
For $n\geq 1$, the {\em boolean $n$-cube\/}  is a graph $C^n=(V,E,\psi)$, where $V=B^n$ (\emph{vertices}), $E=\{e=(v^i,v^j)\mid v^i,v^j\in V \txt{and} v^i \txt{and} v^j \text{ differ in exactly one coordinate 
\hspace{-2.5mm}
}\}$ (\emph{edges}), and $\psi(e)=\psi((v^i,v^j))=(v^i,v^j)$ (\emph{incidence funtion}). 
For a boolean function $f:B^n\to B$, $n\geq 1$, the {\em cube\/} $C^n_f$ of $f$ is the  $n$-cube where $f(v)$ is assigned to each vertex $v\in V=B^n$.
If $n$ is understood, we denote $C^n_f$ by $C_f$.

In a function cube $C_f$, an edge $e=(v^i,v^j)\in E$ is {\em live\/}  if $f(v^i)\not=f(v^j)$; otherwise it is {\em dead.\/} 
A {\em live graph\/} of $f$ is that subgraph $L_f$ of $C_f$ that consists of all the live edges and their incident vertices. 
\newcommand{\dist}{\mathrm{dist}}

Instead of considering paths in a digraph $D_f(\Tx)$, we will examine walks in the cube $C_f$.
The size of a digraph $D_f(\Tx)$ increases as the length of $\Tx$ increases, and the length of any path in $D_f(\Tx)$ increases accordingly. However, the size of a cube $C_f$ is independent of any vector $\Tx$, if the dimension of $\Tx$ is fixed.

\begin{example}
The cube of $f=x_1+x'_2$ is shown in Fig.~\ref{square}~(a), where a vertex is white if $f(v^i)=0$, and black otherwise.
We also write 00 instead of $(0,0)$, \etc, to simplify the notation.
The live edges of $f$ are shown by thick lines in Fig.~\ref{square}~(b). The live graph of $f$   is shown in Fig.~\ref{square}~(c).

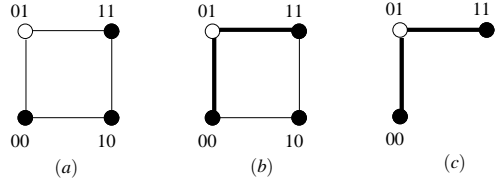
\begin{figure}[hbt]
\begin{center}
\setlength{\unitlength}{0.00050000in}
\begingroup\makeatletter\ifx\SetFigFont\undefined%
\gdef\SetFigFont#1#2#3#4#5{%
  \reset@font\fontsize{#1}{#2pt}%
  \fontfamily{#3}\fontseries{#4}\fontshape{#5}%
  \selectfont}%
\fi\endgroup%
{\renewcommand{\dashlinestretch}{30}
\begin{picture}(5048,1870)(0,-10)
\put(2865,1693){\makebox(0,0)[lb]{\smash{{\SetFigFont{7}{8.4}{\familydefault}{\mddefault}{\updefault}$11$}}}}
\put(1065,643){\blacken\ellipse{150}{150}}
\put(1065,643){\ellipse{150}{150}}
\put(165,643){\blacken\ellipse{150}{150}}
\put(165,643){\ellipse{150}{150}}
\put(1065,1543){\blacken\ellipse{150}{150}}
\put(1065,1543){\ellipse{150}{150}}
\put(4065,1558){\ellipse{150}{150}}
\put(4065,658){\blacken\ellipse{150}{150}}
\put(4065,658){\ellipse{150}{150}}
\put(4965,1558){\blacken\ellipse{150}{150}}
\put(4965,1558){\ellipse{150}{150}}
\put(2115,1543){\ellipse{150}{150}}
\put(2115,643){\blacken\ellipse{150}{150}}
\put(2115,643){\ellipse{150}{150}}
\put(3015,643){\blacken\ellipse{150}{150}}
\put(3015,643){\ellipse{150}{150}}
\put(3015,1543){\blacken\ellipse{150}{150}}
\put(3015,1543){\ellipse{150}{150}}
\path(240,643)(990,643)
\path(1065,1468)(1065,718)
\path(172,1453)(172,710)
\path(247,1550)(990,1550)
\blacken\path(4140,1573)(4890,1573)(4890,1543)
	(4140,1543)(4140,1573)
\path(4140,1573)(4890,1573)(4890,1543)
	(4140,1543)(4140,1573)
\path(2190,643)(2940,643)
\path(3015,1468)(3015,718)
\blacken\path(2190,1573)(2940,1573)(2940,1543)
	(2190,1543)(2190,1573)
\path(2190,1573)(2940,1573)(2940,1543)
	(2190,1543)(2190,1573)
\blacken\path(2145,1468)(2145,718)(2115,718)
	(2115,1468)(2145,1468)
\path(2145,1468)(2145,718)(2115,718)
	(2115,1468)(2145,1468)
\blacken\path(4095,1468)(4095,718)(4065,718)
	(4065,1468)(4095,1468)
\path(4095,1468)(4095,718)(4065,718)
	(4065,1468)(4095,1468)
\put(472,73){\makebox(0,0)[lb]{\smash{{\SetFigFont{7}{8.4}{\familydefault}{\mddefault}{\updefault}$(a)$}}}}
\put(2505,81){\makebox(0,0)[lb]{\smash{{\SetFigFont{7}{8.4}{\familydefault}{\mddefault}{\updefault}$(b)$}}}}
\put(15,343){\makebox(0,0)[lb]{\smash{{\SetFigFont{7}{8.4}{\familydefault}{\mddefault}{\updefault}$00$}}}}
\put(915,343){\makebox(0,0)[lb]{\smash{{\SetFigFont{7}{8.4}{\familydefault}{\mddefault}{\updefault}$10$}}}}
\put(15,1693){\makebox(0,0)[lb]{\smash{{\SetFigFont{7}{8.4}{\familydefault}{\mddefault}{\updefault}$01$}}}}
\put(915,1693){\makebox(0,0)[lb]{\smash{{\SetFigFont{7}{8.4}{\familydefault}{\mddefault}{\updefault}$11$}}}}
\put(3915,358){\makebox(0,0)[lb]{\smash{{\SetFigFont{7}{8.4}{\familydefault}{\mddefault}{\updefault}$00$}}}}
\put(3915,1708){\makebox(0,0)[lb]{\smash{{\SetFigFont{7}{8.4}{\familydefault}{\mddefault}{\updefault}$01$}}}}
\put(4815,1708){\makebox(0,0)[lb]{\smash{{\SetFigFont{7}{8.4}{\familydefault}{\mddefault}{\updefault}$11$}}}}
\put(4515,118){\makebox(0,0)[lb]{\smash{{\SetFigFont{7}{8.4}{\familydefault}{\mddefault}{\updefault}$(c)$}}}}
\put(1965,343){\makebox(0,0)[lb]{\smash{{\SetFigFont{7}{8.4}{\familydefault}{\mddefault}{\updefault}$00$}}}}
\put(2865,343){\makebox(0,0)[lb]{\smash{{\SetFigFont{7}{8.4}{\familydefault}{\mddefault}{\updefault}$10$}}}}
\put(1965,1693){\makebox(0,0)[lb]{\smash{{\SetFigFont{7}{8.4}{\familydefault}{\mddefault}{\updefault}$01$}}}}
\put(165,1543){\ellipse{150}{150}}
\end{picture}
}
\end{center}
\caption{Graphs for $f=x_1+x'_2$: (a) $C_f$; (b) live edges; (c) $L_f$.}
\label{square}
\end{figure}

\end{example}

We extend the concept of cost to paths in digraphs and walks in cubes. For a path $P = \Tv^1,\ldots,\Tv^m$ in $D_f(\Tx)$, let $c_P=|E_=(P)|$, where
$
E_=(P) = \{(\Tv^i,\Tv^{i+1}) \mid \lambda(\Tv^i) = \lambda(\Tv^{i+1}), ~ i = 1,\ldots,m-1\}.
$
Thus $c_P$ is the number of arcs in $P$ whose endpoints have the same $\lambda$ values. 
For any walk $W = w^1,w^2,\cdots,w^m$ in $C_f$, let $c_W=|E_=(W)|$, where 
$
E_=(W) = \{(w^j,w^{j+1}) \mid f(w^j) = f(w^{j+1}),~ j = 1,\ldots,m-1\}.
$
Thus $c_W$ is the number of edges in $W$ whose endpoints have the same $f$ values, that is, the number of edges in $W$ which are not in the live graph~$L_f$.

\begin{definition} \label{def:pathwalks}
Let $f:B^n\to B$ be a boolean function, and $\Tf$, its extension.
Let $\Tx\in \TT^n$ be a  vector, and $P= \Tv^1,\ldots,\Tv^r$,  a  path in $D_f(\Tx)$ from  $\Tv^1=\alpha(\Tx)$ to  $\Tv^r=\Tx$. Let $W(P)$ be the sequence
$W(P)= \omega(\Tv^1),\ldots, \omega(\Tv^r).$
Conversely, let $W=w^1,\ldots, w^r$ 
be any walk in $C_f$, where $w^i\in B^n$, for $i=1,\ldots,r$.  
Let $P(W)$ be the sequence 
$P(W)=w^1,w^1\circ w^2,\ldots,w^1\circ\cdots\circ w^r.$
\end{definition}

\begin{theorem}\label{thm:pathwalks}
If $P$ is a path in $D_f(\Tx)$ then $W(P)$ is a walk in $C_f$ and $c_P = c_{W(P)}$.
If $W=w^1,\ldots, w^r$ is a walk in $C_f$, let $\Tx=w^1\circ\cdots\circ w^r$.
Then $P(W)$ is a path in $D_f(\Tx)$ and $c_W = c_{P(W)}$.
Moreover, if $P$ is a  path in $D_f(\Tx)$, then $P(W(P))=P$, and
if $W$ is a walk in $C_f$, then $W(P(W))=W$.
\end{theorem}
A walk $W$ is {\em optimal\/} if path $P(W)$ is optimal. 

\begin{example} 
In Fig.~\ref{extension},  $P=\Tv^1,\ldots,\Tv^6$ is a  path from $\Tv^1=\alpha(\Tx)$ to $\Tv^6=\Tx$, and
$c_P = |\{(\Tv^4,\Tv^5),(\Tv^5,\Tv^6)\}| = 2$.
Let
$w^i=\omega(\Tv^i)$, for $i=1,\ldots,6$;
then $W(P)=w^1,\ldots,w^6$ is a walk in Fig.~\ref{square} (a). Note that $(w^4,w^5)$ and $(w^5,w^6)$ are not in the live graph $L_f$; thus $c_{W(P)} = 2 = c_P$.

Conversely, for $W=w^1,\ldots,w^6=(01,00,10,11,10,00)$, let $\Tx^1=w^1=(0,1)$,
$\Tx^2=w^1\circ w^2=(0,1)\circ(0,0)=(0\circ 0, 1\circ 0)=(0,10)$, $\Tx^3=(0,101)$, $\Tx^4=(01,101)$, $\Tx^5=(01,1010)$, and $\Tx^6=(010,1010)$. Then $P(W) = \Tx^1,\ldots,\Tx^6$ is a  path in $D_f(\Tx)$, and $c_{P(W)} = 2 = c_W$. In addition, $P(W)=P(W(P))=P$.
Here $c_P\neq 0$ and $W$ is not a walk in $L_f$. 
If $\Tu^1=(0,1)$, $\Tu^2=(01,1)$, $\Tu^3=(010,1)$, $\Tu^4=(010,10)$, $\Tu^5=(010,101)$,  $\Tu^6=(010,1010)$, and $P=\Tu^1\cdots\Tu^6$,  then $c_P=0$, and  $W(P)=01,11,01,00,01,00$ is a walk in $L_f$.
\end{example}

In the rest of the paper we consider walks in the $n$-cube $C_f$. A walk $W=w^1,\ldots, w^r$ is a \emph{walk for a vector $\Tx$} if $\Tx=w^1\circ\cdots\circ w^r$.
To evaluate $\Tf(\Tx)$ we find a walk $W=w^1,\ldots, w^r$ for $\Tx$ with minimal cost, and then  $\Tf(\Tx)=f(w^1)\circ \cdots\circ f(w^r)$. This approach takes the advantage of the fact that the size of the $n$-cube $C_f$ is independent of the length $l(\Tx)$ of the input vector $\Tx$.

\section{Characteristic vectors}
\label{sec:char}
Now we are interested only in proper vectors. A vector is \emph{minimal} if each component has length 2 or 3.
 If a transient $\Tt=t_1\cdots t_m$ is proper,  the {\em characteristic transient} of $\Tt$ is $\tilde{\Tt}$, where
 $\tilde{\Tt} =  t_1t_2$ if $m$ is even, and $\tilde{\Tt} =  t_1t_2t_3$ if $m$ is odd.
Note that $\tilde{\Tt}$ is a prefix of $\Tt$, $\alpha(\tilde{\Tt}) = \alpha(\Tt)$,  $\omega(\tilde{\Tt}) = \omega(\Tt)$, and $\Delta(\Tt) \equiv \Delta(\tilde{\Tt})$, where $\equiv$ is equivalence modulo 2.
If $\Tx = (\Tx_1,\ldots,\Tx_n)$ is a proper vector, then $\tilde{\Tx} = (\tilde{\Tx}_1,\ldots,\tilde{\Tx}_n)$ is the {\em characteristic vector} of $\Tx$.
Also,
$\tilde{\Tx}$ is a prefix of~ $\Tx$,
$\alpha(\tilde{\Tx}) = \alpha(\Tx)$, $\omega(\tilde{\Tx}) = \omega(\Tx)$,
and $\Delta(\Tx_i) \equiv \Delta(\tilde{\Tx_i})$, for $i\in[n]$.  The characteristic vector of any vector is minimal, and every minimal vector is the characteristic vector of some vector.
Any vector $\Tx$ which has $\tilde{\Tx}$ as its characteristic vector is an \emph{prolongation} of $\tilde{\Tx}$.

A function $f : B^n \rightarrow B$ {\em depends} on its $k$-th argument if there exist $x_i \in B$, such that 
$$f(x_1, \ldots, x_{k-1}, 0, x_{k+1}, \ldots, x_n) \neq f(x_1, \ldots, x_{k-1}, 1, x_{k+1}, \ldots, x_n).$$
In this section, we only consider functions that depend on all of their arguments. 
If $f : B^n \rightarrow B$ depends on $x_k$,  then there exists at least one live edge $e = (w^i,w^j)$ in the cube $C_f$ of $f$,  where $w^i$ and $w^j$ differ only in $x_k$.

\begin{definition}
\label{def:convenient}
A  boolean function $f : B^n \rightarrow B$ that depends on all of its variables is \emph{convenient} if, for every proper vector 
$\Tx$, the cost $c_f(\Tx)$ of  $\Tx$ is equal to the cost $c_f(\tilde{\Tx})$ of its characteristic vector~$\tilde{\Tx}$; otherwise, $f$ is \emph{inconvenient}. 
\end{definition}

For any vector $\Tx \in \TT^n$, let $\tilde{\varphi}(\Tx) = \varphi(\tilde{\Tx})$ be the number of walks for $\tilde{\Tx}$. 
When $\Delta(\tilde{\Tx}_1) = \cdots = \Delta(\tilde{\Tx}_n) = 2$, the maximal value of $\tilde{\varphi}(\Tx)$ is obtained from Equation~(\ref{eq:phi}) by setting $m=2n$, and is approximately 
\begin{equation}
\tilde{\varphi}(\Tx) \approx (4\pi)^{\frac{1-n}{2}}n^{2n+\frac{1}{2}},
\end{equation}
which is independent of the length $m$ of $\Tx$; hence the evaluation of a particular function $\Tf$ can be much more efficient if $f$ is convenient. We first search for an optimal walk for $\tilde{\Tx}$, and get its cost $c$; we then compute the boolean value $f(\alpha(\Tx))$ and construct a transient  with $\Delta(\Tx) - c$ changes beginning with $f(\alpha(\Tx))$. With fixed $n$, this can be done in time linear in the length of $\Tx$.

The next claim follows immediately from Theorem~\ref{thm:costg}.

\begin{corollary}
All functions in $\cal G$ are convenient. 
\end{corollary}

To prove that a function $f$ is convenient we must verify that, for every vector $\Tx$,
the cost of $\Tx$ for $f$ is the same as the cost of $\tilde{\Tx}$ for $f$. 
Equivalently, we need to show that, for every minimal vector $\tilde{\Tx}$, the cost for $f$ of any prolongation $\Tx$ of $\tilde{\Tx}$ is the same as the cost of $\tilde{\Tx}$ for $f$.

An edge $e$ in a cube corresponds to the unique coordinate $x_i$ that has complementary values in the two vertices of that edge; we say that $e$ is \emph{an edge in  coordinate $x_i$}.
An edge is {\em incident} to a walk $W$ if it shares at least one vertex with $W$.
A walk $W$ is {\em complete} if, for every coordinate, there is a live edge in that coordinate incident to $W$. 
A vertex $v$ in a cube $C_f$ of a boolean function $f$ is a {\em focus\/} if every edge incident to $v$ is live. Any walk through a focus is complete. 

 \begin{prop}
 \label{prop:complete}
Let $f : B^n \rightarrow B$ be a boolean function, and let $C_f$ be its cube. Let $\Tx$ be a transient vector and   $\tilde{\Tx}$  its characteristic vector.
If an optimal walk $W$  in $C_f$ for $\tilde{\Tx}$ is complete, then the cost of any optimal walk for $\Tx$ is $c_W$.
 \end{prop}

\noin
{\it Proof:\/}
Let $W$ be an optimal walk for $\tilde{\Tx}$. Since the difference between the number of changes in $\Tx_i$ and in $\tilde{\Tx}_i$ is even for any $i$, the additional changes in $\Tx_i$ can be inserted after a vertex incident to the live edge in that coordinate is reached.

\section{3-variable functions}
\label{sec:sym}
Suppose $f:B^n\to B$ and $g:B^n\to B$ are boolean functions. If $g(y_1,\ldots,y_n)$ 
can be obtained from $f(x_1,\ldots,x_n)$ by renaming the variables and complementing some number of inputs and/or the output, then we write $f\sim g$, where  $\sim$ is an equivalence 
relation~\cite{Cal58,Har65}, and we say that $f$ and $g$ are in the same \emph{symmetry class}. 
For example, we can start with $x+y$, rename $y$ as $z$ to get $x+z$, complement $z$ to get 
$x+z'$, and complement this result to get $x'z$.
Thus $x'z \sim x+y$.
If we know how to evaluate the transient extension of $f$, then we can also evaluate the transient extensions of all the functions that are in the same symmetry class as $f$~\cite{BrYe10}. Hence we consider only one representative function of each symmetry class.

\begin{figure}[ht]
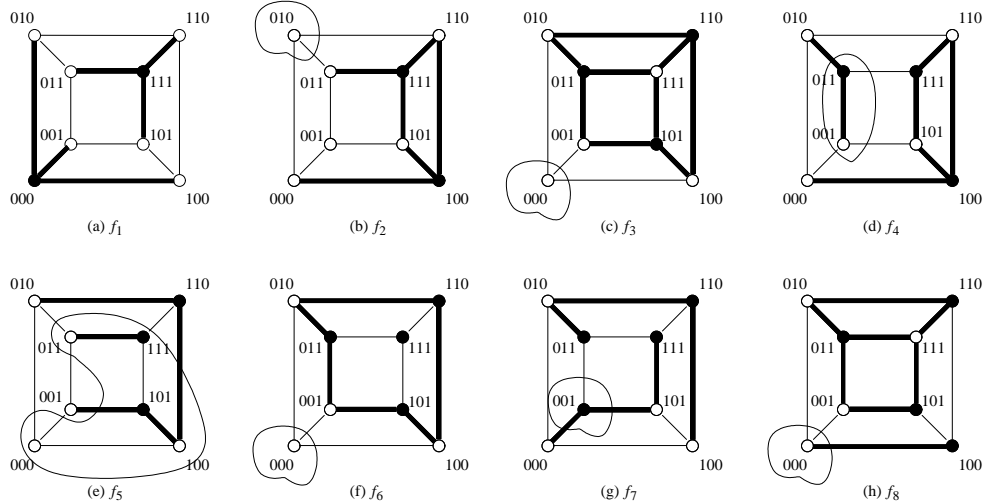

\begin{center}
 \input reduced_cube.eepic
\end{center}
\caption{Representatives of the eight symmetry classes of 3-variable functions.}
\label{cube3var}
\end{figure}

For $n = 3$, there are 256 functions which can be reduced to 14 symmetry classes~\cite{Cal58,Har65}. Four  classes, represented by $0$, $x$, $x+y$, and $x\oplus y$, contain degenerate functions; these classes account for 38 functions which can all be evaluated
using the formulas in Section~\ref{sec:trans}. The  remaining 218 functions can be reduced to 2 symmetry classes (18 functions) in $\cG$  and  8  classes (200 functions) represented by the functions shown in Fig.~\ref{cube3var}, where  the circled vertices can be ignored for now. 

The main result in this section is the following:

\begin{theorem}
\label{thm:3var}
All 3-variable functions are convenient.
\end{theorem}

To prove Theorem~\ref{thm:3var}, it is sufficient to examine the  eight functions of Fig.~\ref{cube3var}.
The following results are useful in proving that certain walks are optimal: 
\begin{lemma}
\label{lem:off1}
Let $W_1$ and $W_2$ be two walks from vertex $u$ to vertex $v$. Then the difference between the cost of $W_1$ and that of $W_2$ is a multiple of 2.
\end{lemma}

%


\begin{corollary}
\label{cor:cost}
Let $c\ge0$ be any integer. If there is no walk from $u$ to $v$ of cost less than or equal to $c-1$, and there is a walk $W$ of cost $c+1$, then $W$ is optimal.
In particular, every walk of cost 1 from $u$ to $v$ is optimal.
\end{corollary}

\begin{lemma}
\label{lem:specialv}
For a 3-variable function and a minimal vector $\Tx$, if  $\alpha(\Tx)$ is  within distance 1 from a focus, 
then there is a  complete optimal walk for~$\Tx$.
\end{lemma}

For any 3-variable function $f$, a walk $v_0v_1v_2v_3v_4v_5$ on $C_f$ is \emph{alternating} if each subwalk $v_iv_{i+1}v_{i+2}v_{i+3}$, $i=0, 1, 2$, contains a change in every coordinate.

\begin{lemma}
\label{lem:len5}
Let $\Tx$ be a minimal vector for a 3-variable function,  let $U=v_0v_1v_2v_3$  be any walk such that  $v_0=\alpha(\Tx)$, and  let $W=Uv_4v_5$.
If $W$ is alternating, $c(U)=0$, and $c(W)\le 1$, then there exists a complete optimal walk $V$ for $\Tx$.
\end{lemma}

We are now ready to sketch a proof of Theorem~\ref{thm:3var}.

\noin
{\it Proof:\/}
For each of the eight functions in Fig.~\ref{cube3var}, we enumerate all minimal vectors 
$\Tx=\langle\alpha(\Tx); \Delta_1,\Delta_2,\Delta_3\rangle$, and prove that there is a complete optimal walk for each $\Tx$. 
The vertices that need to be considered are circled in the figure.
Since there are eight  functions and each of them has eight possible starting vertices $\alpha(\Tx)$ and eight change vectors $(\Delta_1,\Delta_2,\Delta_3)$,
there are 512 cases to analyze. We reduce this number significantly by using Corollary~\ref{cor:cost}, Lemmas~\ref{lem:specialv} and~\ref{lem:len5}, and symmetry. 
The value $4b_1+2b_2+b_3$  represents vertex $(b_1, b_2, b_3)$, and
we denote walks by words, rather than sequences. The eight functions are treated as follows:
\smallskip

\noin
1. For $f_1$, every vertex is within distance 1 from a focus; by Lemma~\ref{lem:specialv}, no vertex needs to be considered. 
\smallskip

\noin
2. For $f_2=x_1(x_2\oplus x_3)$, only vertices $1$ and $2$ are not within distance 1 of a focus. Since $f_2$ is symmetric in $x_2$ and $x_3$, we consider only one of 1 and 2, say~$2$. We list the complete walks in pairs $(\Delta_1\Delta_2\Delta_3,W)$, where $W$ is a complete optimal walk for 
$\langle2; \Delta_1,\Delta_2,\Delta_3\rangle$. There are no minimal walks of cost 0.  The  walks of cost 1 are: 
$(111,2645)$, $(112, 26454)$, $(121,26467)$, $(122,264676)$, $(212,267640)$, and $(221,264673)$. 
Walks $(211,26451)$ and $(222,2646762)$ are of cost 2; by Corollary~\ref{cor:cost}, they are optimal.
Each walk is complete for  it goes through a focus.
\smallskip

\noin
3. For $f_3$, only $0$ is not within distance 1 of a focus, so we consider it.
\smallskip

\noin
4. For $f_4=x_2x_3+x_1x_2'x_3'$,  $1,2,3,7$  are not within distance 1 of a focus. 
Since  $f_4$ is symmetric in $x_2$ and $x_3$, we consider only 1 and not 2.
For  7, there is a walk $W=754023$, which satisfies the conditions of Lemma~\ref{lem:len5}.
So  there is a complete optimal walk for any minimal vector starting at 7, and
 we consider only 1  and 3.
 \smallskip

\noin
5. The function  $f_5=x_1(x_2+x_3)$ has no focus. Since it is symmetric in $x_2$ and $x_3$, we  consider only 1 (and not 2) and 5 (and not 6), say.  Walk 154623 satisfies the conditions of Lemma~\ref{lem:len5}, taking care of 1. So we consider  0, 3, 4, 5, and 7.
\smallskip

\noin
6. For $f_6=x_1x_2+x_2x_3+x_3x_1$, there is no focus. Because $f$ is symmetric in all three variables, it suffices to consider 1 (and not 2 and 4), 6 (and not 3 and 5), 0 and 7.
Moreover, if we complement the function, the live and dead edges are preserved. Hence 1 and 6 are symmetric in the cube as are 0 and 7, and we consider only 0 and 1. However, for vertex 1,  walk $132645$ meets the conditions of Lemma~\ref{lem:len5}. Hence we  look only at 0.
\smallskip

\noin
7. For  $f_7$,
vertices 0, 3, 4, and 7 are symmetric in the live graph, as are 1, 2, 5, and 6. The alternating walk $015762$ takes care of $0$, leaving only 1 to consider.
\smallskip

\noin
8. For $f_8$,
only $0$ and $4$ are not within distance 1 of a focus, and they are symmetric with respect to live edges. So we examine only 0.
\smallskip

\noin

We now show a function which is not convenient. Let $S_{i}(x_1,x_2,x_3,x_4)$ be the symmetric function of four variables that is 1 if and only if precisely $i$ of its variables are 1. 
Also, let $S_{2,3}(x_1,x_2,x_3,x_4)=S_2(x_1,x_2,x_3,x_4)+S_3(x_1,x_2,x_3,x_4)$.
\begin{prop}
\label{prop:inconvenient} 
$f=S_{2,3}(x_1,x_2,x_3,x_4)
+x_0x_1x_2x_3x_4
$
is inconvenient.
 \end{prop}
 

\section{Conclusions}
\label{sec:conc}

The evaluation of extensions of boolean functions is  simplified if we use walks in boolean cubes instead of paths in digraphs. The evaluation of extensions of convenient functions can be done in polynomial time if we use characteristic vectors. 
All 3-variable functions are convenient, but there exist inconvenient 5-variable functions. It remains open whether there is an inconvenient 4-variable function. 
The problem of characterizing convenient functions is also open. 


\bibliographystyle{eptcs}
\bibliography{2010_DCFS_BrzozowskiLiYe}

\end{document}